\documentclass[10pt,english,aps,amssymb,prb,twocolumn, showpacs]{revtex4-1}
\pdfoutput=1
\usepackage[T1]{fontenc}
\usepackage[latin9]{inputenc}
\usepackage{amsmath}
\usepackage{graphicx}
\usepackage{amssymb}
\usepackage{mathtools}
\usepackage{esint}
\usepackage{verbatim}
\usepackage{bm}
\usepackage{hyperref}

\makeatletter
\@ifundefined{textcolor}{}
{
 \definecolor{WHITE}{gray}{1}
 \definecolor{RED}{rgb}{1,0,0}
 \definecolor{GREEN}{rgb}{0,1,0}
 \definecolor{BLUE}{rgb}{0,0,1}
 \definecolor{CYAN}{cmyk}{1,0,0,0}
 \definecolor{MAGENTA}{cmyk}{0,1,0,0}
 \definecolor{YELLOW}{cmyk}{0,0,1,0}
 }

\renewcommand{\phi}{\varphi}
\renewcommand{\epsilon}{\varepsilon}

\renewcommand{\vec}[1]{{\bf #1}}
\renewcommand{\vr}[1]{{\bf #1}}

\usepackage{babel}

\begin{document}

\title {Chern mosaic -- topology of chiral superconductivity on ferromagnetic adatom lattices}
\author{Joel R\"ontynen}
\email[Correspondence to ]{joel.rontynen@aalto.fi}
\author{Teemu Ojanen}
\email[Correspondence to ]{teemuo@boojum.hut.fi}
\affiliation{Department of Applied Physics (LTL), Aalto University, P.~O.~Box 15100,
FI-00076 AALTO, Finland }
\date{\today}
\begin{abstract}
In this work we will explore the properties of superconducting surfaces decorated by two-dimensional ferromagnetic adatom lattices. As discovered recently [R\"ontynen and Ojanen, Phys. Rev. Lett. \textbf{114}, 236803 (2015)], in the presence of a Rashba spin-orbit coupling these systems may support topological superconductivity with complex phase diagrams and high Chern numbers. We show how the long-range hopping nature of the effective low-energy theory generically gives rise to a phase diagram covered by a \emph{Chern mosaic} -- a rich pattern of topological phases with large Chern numbers. We study different lattice geometries and the dependence of energy gaps and abundance of different phases as a function of system parameters. Our findings establish the studied system as one of the richests platforms for topological matter known to date.      
       
\end{abstract}
\pacs{73.63.Nm,74.50.+r,74.78.Na,74.78.Fk}
\maketitle
\bigskip{}

\section{introduction}

The search for novel solid-state phases is advancing on various fronts. It has recently become clear that viable paths to novel topological states of matter\cite{volovik} increasingly employ engineering suitable conditions and hybrid structures instead of simply relying on cataloguing material properties. While finding materials that exhibit intrinsic topological superconductivity is in principle possible, it could be more convenient to combine established materials, structures and techniques to achieve the desired outcome. 

The experimental observation of signatures of Majorana bound states in a ferromagnetic chain of iron atoms deposited on a superconducting surface represents the latest promising development in the pursuit for topological superconductivity.\cite{np2} The recent experiment is an important addition to the previous successes\cite{mourik, das} in nanowire hybrid structures which display intriguing signatures of topological superconductivity. These efforts can be regarded as a testament to the power of artificially engineering novel quantum states of matter. Ideally these developments will lead, among other things, to the realization of non-abelian quasiparticles with applications in quantum information.\cite{nayak} 

An interplay of magnetic adatoms and superconductivity offers interesting possibilities in engineering topological superconductors starting from a trivial state.\cite{choy, np,poyh, pientka2, pientka3} Magnetic atoms in an $s$-wave superconductor give rise to bound states inside the superconducting gap.\cite{yu, shiba, rusinov, balatsky, yaz} In a regular lattice of magnetic atoms deposited sufficiently close to each other the bound states of nearby atoms hybrize and form subgap energy bands.\cite{bry, heimes, heimes2, ront, vazifeh, reis,west, poyh2, zhang, schecter}  Recent theoretical studies have revealed in detail how the 1d magnetic atom chain may give rise to topological superconductivity as in Kitaev's $p$-wave chain.\cite{kitaev1} The topological properties of the chain depend on the magnetic texture of the lattice, the distance of the atoms, the spin-orbit coupling in the superconducting surface, the strength of the magnetic moments and a number of other microscopic details.\cite{brau1, brau2, klin, heimes2, vazifeh, schecter} These parameters lead to rich variation of possible topological states in the chain. 

\begin{figure}
\includegraphics[width=0.99\columnwidth]{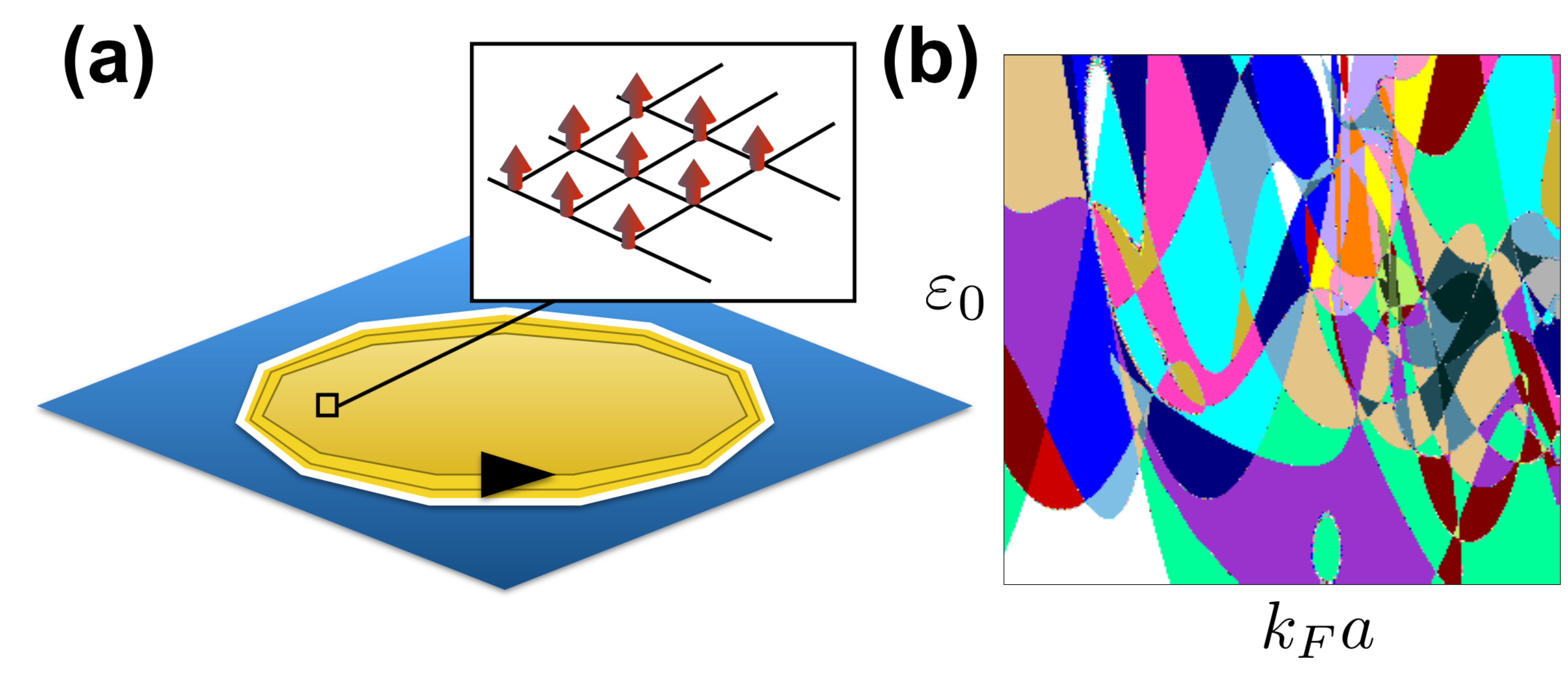} 
\caption{(a) Studied system consists of a lattice of magnetic atoms on a superconducting film. The area covered by magnetic atoms supports topological superconductivity enclosed by circulating edge modes. (b) Chern mosaic- a topological phase diagram of the studied system as a function of on-site bound-state energy $\epsilon_0$ and the effective hybridization parameter $k_Fa$. Different colours correspond to different Chern numbers. }
\label{first}
\end{figure}
Motivated by the recent experimental developments,\cite{np2, pawlak} it was discovered that 2d arrays of magnetic adatoms on an $s$-wave superconductor realizes topologically nontrivial chiral superconductivity.\cite{ront1, li1, nakosai} These systems are illustrated  in Fig.~\ref{first} (a).  In the case of a dilute adatom lattice these systems have striking properties such as topological phases with large Chern numbers of the order of $\xi/a$, where $\xi$ is the superconducting coherence length and $a$ is the adatom lattice constant.\cite{ront1} The  topological phase diagram can, for example, look like the one depicted in Fig.~\ref{first} (b) as a function of on-site bound state energy $\epsilon_0$ and the effective hopping parameter $k_Fa$, where $k_F$ denotes the Fermi wavenumber of the underlying superconductor. Since the fraction may realistically take values of the order of $\xi/a\sim 1-100$, it is clear that the system exhibits one of the most complicated phase diagrams of any physically motivated system.\cite{ront1} The different states labelled with Chern number $C$ are topologically equivalent to chiral superconductors with $(p_x\pm ip_y)^{|C|}$ pairing, where the sign (chirality) is determined by the sign of $C$. In superconducting systems, the Chern number determines the number of chiral Majorana edge modes that support quantized heat conductance in the system. Thus high Chern numbers are preferable in achieving more efficient edge channel transport. Signatures of these edge modes could be observed by Scanning Tunneling Microsopy (STM) in a similar fashion to how Majorana bound states are probed in 1d chains.\cite{ront1}  

In this work, we will build on the previous discovery of topological superconductivity in a ferromagnetic square lattice\cite{ront1} and study the complex competition of different topological phases in more detail. Topological properties in different lattice geometries generically exhibit a similar fascinating structure which we will call a Chern mosaic. The complexity of the phase diagrams is well-hidden in the analytical form of the effective low-energy theory and is revealed only by numerical evaluation of topological invariants and energy gaps. In Sec.~\ref{ferro} we will introduce the studied model and derive an effective tight-binding Hamiltonian describing arbitrary collections of magnetic moments on a 2d superconductor with a Rashba coupling. Then we will focus on ferromagnetic textures, derive momentum space Bogoliubov-de Gennes Hamiltonians for rectangular, triangular and honeycomb lattices, and give the corresponding expressions for their Chern numbers. Topological phase diagrams for various lattices are presented in Sec.~\ref{topophase}, and in Sec.~\ref{mosaic} we examine the Chern mosaic properties as a function of system parameters. In Sec.~\ref{summary} we will discuss and summarize our findings.       

\section{Theoretical formulation}\label{ferro}

\subsection{Effective low-energy model}

In this section, we will formulate the theory describing the systems in Fig.~\ref{first} and introduce theoretical tools and concepts that will be employed in the paper. The bulk electrons in a 2d superconductor are described by a Bogoliubov-de Gennes Hamiltonian
\begin{equation*}
\mathcal{H}_\vec{k}^{(\rm bulk)} = \tau_z \big[ \xi_\vec{k} \sigma_0 + \alpha_{\rm R}( k_y\sigma_x - k_x\sigma_y )\big] + \Delta\tau_x\sigma_0,
\end{equation*}
in the Nambu basis $(\hat{\Psi}_{\uparrow}, \hat{\Psi}_{\downarrow}, \hat{\Psi}^\dagger_{\downarrow}, -\hat{\Psi}^\dagger_{\uparrow})$. Here $\xi_\vr{k} = \frac{\hbar^2k^2}{2m} - \mu$ with the Fermi energy $\mu$, $\alpha_{\rm R}$ is the Rashba spin-orbit coupling (SOC) strength and $\Delta$ is the superconducting pairing amplitude. The two set of Pauli matrices $\tau_i$ and $\sigma_i$ operate in the particle-hole and spin space and $\sigma_0$ denotes $2\times2$ unit matrix. The electrons interact with the magnetic adatoms by an exchange interaction of strength $J$:
\begin{equation*}
\mathcal{H}^{(\rm imp)}(\vec{r}) = -J\sum_n \vec{S}_n\cdot\bm{\sigma}\,\delta(\vec{r}-\vec{r}_n),
\end{equation*}
where $\vr{r}_n$ are the positions of the atoms and $\vr{S}_n$ the corresponding spins. The total Hamiltonian is a combination of the two terms, $\mathcal{H} = \mathcal{H}^{(\rm bulk)} + \mathcal{H}^{(\rm imp)}$.

It has been known since the work of Yu, Shiba and Rusinov\cite{yu, shiba, rusinov} that a single magnetic impurity binds one fermionic state within the superconducting gap. To simplify the theoretical description, we will consider the limit of deep impurities, $|1-\alpha|\ll 1$, where $\alpha = \pi JS\mathcal{N}$ is a dimensionless impurity strength and $\mathcal{N}$ is the spin-averaged density of states at the Fermi level.  Physically this assumption means that the energy of an isolated impurity state $\epsilon_0=\Delta(1-\alpha)$ lies close to the center of the superconducting gap. We also assume that the impurity separation $a$ is large enough, $k_Fa\gg1$, for the impurity band to be well within the superconducting gap. In the deep-dilute limit, the standard steps\cite{pientka2, pientka3, bry, ront1} yield an effective low-energy tight-binding Hamiltonian\cite{ront1}
\begin{equation}\label{h1}
\begin{gathered}
H_{mn} = \begin{pmatrix} h_{mn} & \Delta_{mn} \\ (\Delta_{mn})^\dagger & -h_{mn}^* \end{pmatrix} ,
\end{gathered}
\end{equation}
where we have projected to the electron- and holelike components of the subgap Shiba states. The effective Hamiltonian (\ref{h1}) has a Bogoliubov-de Gennes block structure of $N\times N$ blocks, where $N$ is the number of magnetic atoms.  The blocks are given by
\begin{widetext}
\begin{equation}
\begin{split}
h_{mn} &= \begin{dcases} \epsilon_0 & m=n \\
\frac{\Delta}{2} \big[ I_1^-(r_{mn}) + I_1^+(r_{mn}) \big] \langle\uparrow_m|\uparrow_n\rangle
+ i \frac{\Delta}{2} \big[ I_3^-(r_{mn}) - I_3^+(r_{mn}) \big] \Big[ \langle\uparrow_m|\sigma_x|\uparrow_n\rangle \frac{y_{mn}}{r_{mn}} - \langle\uparrow_m|\sigma_y|\uparrow_n\rangle \frac{x_{mn}}{r_{mn}} \Big] & m\neq n \end{dcases}
\\
\Delta_{mn} &= \begin{dcases} 0 & m=n\\ 
- \frac{\Delta}{2} \big[ I_2^-(r_{mn}) + I_2^+(r_{mn}) \big] \langle\uparrow_m|\downarrow_n\rangle
- i \frac{\Delta}{2} \big[ I_4^-(r_{mn}) - I_4^+(r_{mn}) \big] \Big[ \langle\uparrow_m|\sigma_x|\downarrow_n\rangle \frac{y_{mn}}{r_{mn}} - \langle\uparrow_m|\sigma_y|\downarrow_n\rangle \frac{x_{mn}}{r_{mn}} \Big] & m\neq n
\end{dcases}
\end{split}
\label{hmn_Dmn_general}
\end{equation}
\end{widetext}
where $r_{mn} = |\vr{r}_m-\vr{r}_n|$ is the distance between two Shiba lattice sites and $x_{mn} = x_m - x_n$, $y_{mn} = y_m - y_n$. Spin states $|\uparrow_n\rangle$ and $|\downarrow_n\rangle$ denote the eigenstates of the local magnetic moments, $\vr{S}_n\cdot\bm{\sigma} |\uparrow_n/\downarrow_n\rangle = \pm |\vr{S}_n| |\uparrow_n/\downarrow_n\rangle$, and the onsite term $\epsilon_0$ stems from the decoupled impurity energy. The matrix elements depend on the functions
\begin{equation*}
\begin{split}
I_1^\pm(r) &= - \frac{\mathcal{N}_\pm}{\mathcal{N}} \text{Re} \Big[J_0(k_F^\pm r + ir/\xi) + iH_0(k_F^\pm r + ir/\xi) \Big],
\\
I_2^\pm(r) &= \frac{\mathcal{N}_\pm}{\mathcal{N}} \text{Im} \Big[J_0(k_F^\pm r + ir/\xi) + iH_0(k_F^\pm r + ir/\xi)\Big],
\\
I_3^\pm(r) &=  \frac{\mathcal{N}_\pm}{\mathcal{N}} \text{Im} \Big[ iJ_1(k_F^\pm r + ir/\xi) + H_{-1}(k_F^\pm r + ir/\xi) \Big] ,
\\
I_4^\pm(r) &= \frac{\mathcal{N}_\pm}{\mathcal{N}} \text{Re} \Big[ iJ_1(k_F^\pm r + ir/\xi) + H_{-1}(k_F^\pm r + ir/\xi) \Big] ,
\end{split}
\end{equation*}
where $J_n$ and $H_n$ are Bessel and Struve functions of order $n$. The Rashba spin-orbit coupling induces two helical bands with respective density of states $\mathcal{N}_\pm=\mathcal{N}(1\mp\lambda/\sqrt{1+\lambda^2})$ and Fermi wavenumber $k_F^\pm = k_F(\sqrt{1 + \lambda^2} \mp \lambda)$, where $k_F$ is the Fermi wavenumber without the spin-orbit coupling, $\lambda = \alpha_R/(\hbar v_F)$ is the dimensionless Rashba coupling and $v_F$ the Fermi velocity in the absence of the spin-orbit coupling. The Rashba coupling also slightly modifies the superconducting coherence length $\xi = \hbar v_F/\Delta \times \sqrt{1 + \lambda^2}$.

For the rest of the article we are considering a ferromagnetic spin texture, $\vr{S}_n = S\hat{e}_z$. Equation  (\ref{hmn_Dmn_general}) then simplifies to
\begin{equation}\label{h2}
\begin{split}
h_{mn} &= \begin{dcases} \epsilon_0 & m=n \\
\frac{\Delta}{2} \Big[ I_1^-(r_{mn}) + I_1^+(r_{mn}) \Big] & m\neq n \end{dcases}
\\
\Delta_{mn} &= \begin{dcases} 0 & m=n \\
\frac{\Delta}{2} \Big[ I_4^-(r_{mn}) - I_4^+(r_{mn}) \Big] \frac{x_{mn} - i y_{mn}}{r_{mn}} & m\neq n. \end{dcases}
\end{split}
\end{equation}
The Hamiltonian ($\ref{h1}$) with entries ($\ref{h2}$) defines an long-range hopping model where the hopping amplitudes decay as $1/\sqrt{r_{mn}}$ at short distances and exponentially at distances $r_{mn}>\xi$. 

\subsection{Momentum space topology}
In order to study the bulk properties we consider an infinite lattice of magnetic moments and pass to momentum space. For any Bravais lattice we can define the following Fourier transforms
\begin{equation*}
\begin{split}
d_x(\vr{k}) &= \text{Re} \sum_{\vr{R}} e^{-i\vr{k}\cdot\vr{R}} \Delta_{\vr{R}},
\\
d_y(\vr{k}) &= - \text{Im} \sum_{\vr{R}} e^{-i\vr{k}\cdot\vr{R}} \Delta_{\vr{R}},
\\
d_z(\vr{k}) &= \sum_{\vr{R}} e^{-i\vr{k}\cdot\vr{R}} h_{\vr{R}},
\end{split}
\end{equation*}
where the sum is over all the lattice vectors $\vr{R}=(x_{mn}, y_{mn})$. The Hamiltonian can then be expressed as $H(\vr{k}) = \vr{d}(\vr{k})\cdot\bm{\sigma}$ with energies $E(\vr{k}) = \pm |\vr{d}(\vr{k})|$. The effective Hamiltonian $H(\vr{k})$ generally defines gapped band structures and satisfies particle-hole symmetry $\mathcal{C}H(k)^*\mathcal{C}^{-1}=-H(-k)$, where $\mathcal{C}=\sigma_x\mathcal{K}$ and $\mathcal{K}$ denotes complex conjugation. Thus the model belongs to the Altland-Zirnbauer class $D$ and its phases are classified by Chern numbers.\cite{schnyder}  The Chern number is given by the expression 
\begin{equation}\label{C1}
C = \frac{1}{4\pi} \smashoperator{\int_{\rm BZ}}\! d^2k\, \frac{\vr{d}}{|\vr{d}|^3} \cdot \bigg( \frac{\partial\vr{d}}{\partial k_1} \times \frac{\partial\vr{d}}{\partial k_2} \bigg).
\end{equation}

In a crystal lattice with a basis of two sites, as in the case of a honeycomb lattice considered below, the momentum space Hamiltonian is a $4\times4$-matrix,
\begin{equation*}
H(\vr{k}) =
\begin{pmatrix}
H_a(\vr{k}) & H_{ab}(\vr{k}) \\ 
H_{ab}(\vr{k})^\dagger & H_b(\vr{k}) 
\end{pmatrix} ,
\end{equation*}
given in the basis $[\hat{a}(\vr{k}), \hat{a}^\dagger(-\vr{k}), \hat{b}(\vr{k}), \hat{b}^\dagger(-\vr{k})]^{\rm T}$ where $\hat{a}(\vr{k})$ and $\hat{b}(\vr{k})$ are the annihilation operators for the two sublattices. The matrix elements are given by
\begin{equation*}
\begin{split}
H_a(\vr{k}) &=H_b(\vr{k}) =
\begin{pmatrix}
h_a(\vr{k}) & \Delta_a(\vr{k}) \\ 
\Delta_a(\vr{k})^* & -h_a(\vr{k}) \\ 
\end{pmatrix} ,
\end{split}
\end{equation*}
\begin{equation*}
\begin{split}
H_{ab}(\vr{k}) &=
\begin{pmatrix}
h_{ab}(\vr{k}) & \Delta_{ab}(\vr{k}) \\ 
-\Delta_{ab}(-\vr{k})^* & -h_{ab}(\vr{k}) \\ 
\end{pmatrix} 
\end{split}
\end{equation*}
in terms of the Fourier transforms
\begin{equation*}
\begin{split}
h_a(\vr{k}) &= \sum_{\vr{R}} e^{-i\vr{k}\cdot\vr{R}} h_{\vr{R}} 
,\\
h_{ab}(\vr{k}) &= \sum_{\vr{R}} e^{-i\vr{k}\cdot(\vr{R}-\vr{d})} h_{\vr{R}-\vr{d}}
,\\
\Delta_a(\vr{k}) &= \sum_{\vr{R}} e^{-i\vr{k}\cdot\vr{R}} \Delta_{\vr{R}} 
,\\
\Delta_{ab}(\vr{k}) &= \sum_{\vr{R}} e^{-i\vr{k}\cdot(\vr{R}-\vr{d})} \Delta_{\vr{R}-\vr{d}}
\end{split}
\end{equation*}
Here we sum over one sublattice and $\vr{d}$ is the vector separating the two basis sites. The subscript $a$ refers to hopping within the same sublattice and $ab$ between the two different sublattices. The Fourier components satisfy the symmetries 
\begin{equation*}
\begin{split}
h_a(-\vr{k}) &= h_a(\vr{k}),
\\
h_{ab}(-\vr{k}) &= h_{ab}(\vr{k})^*,
\\
\Delta_a(-\vr{k}) &= -\Delta_a(\vr{k}).
\end{split}
\end{equation*}
The symmetry of $\Delta_a(\vr{k})$ shows that the effective low-energy model describes an odd-pairing superconductivity. The same conclusion obviously holds for the two-band models determined by the $\vr{d}(\vr{k})$ vector.   

\begin{figure*}
\includegraphics[width=\textwidth]{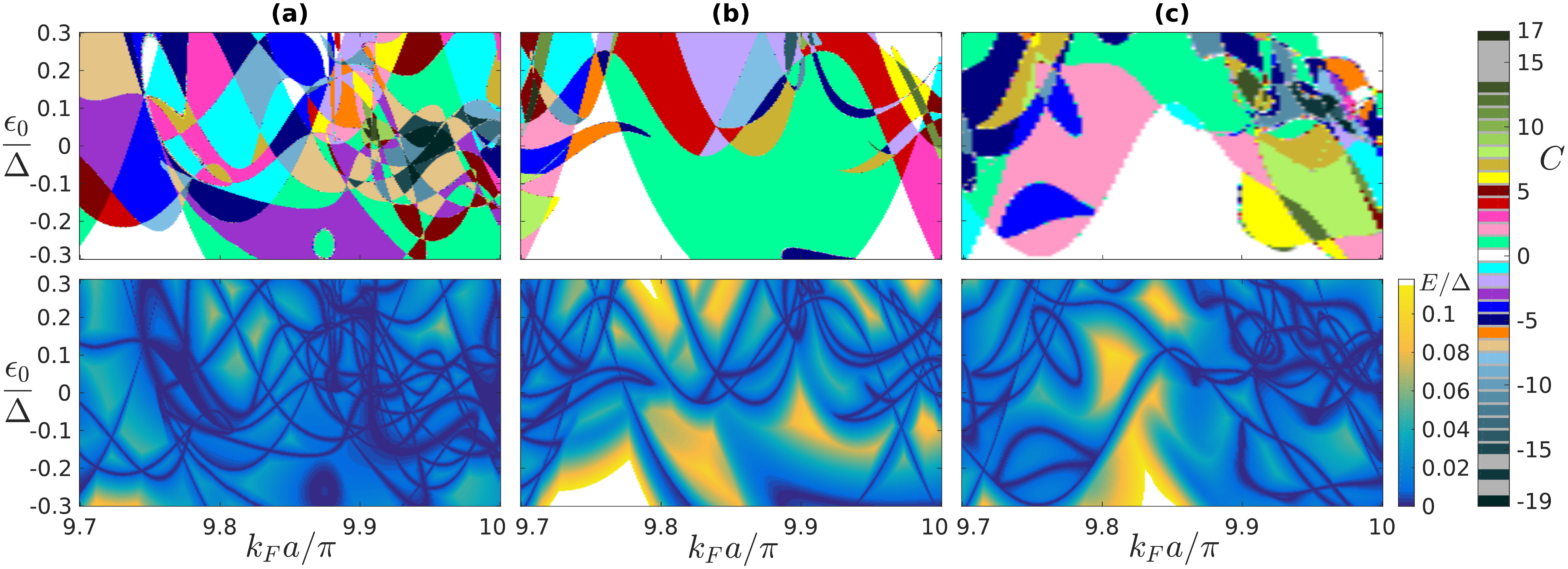} 
\caption{Phase diagrams for various ferromagnetic Shiba lattice geometries. The
Chern number (upper panel) and the corresponding energy gap (lower panel)
in terms of $k_Fa$ and the uncoupled Shiba state energies $\epsilon_0$. We have
considered (a) square, (b) triangular, and (c) honeycomb lattices. Different
topological phases are classified by the Chern number and a topological phase
transition is always accompanied by an energy gap closing. In all figures the
superconducting coherence length $\xi/a=5$ and the dimensionless Rashba coupling
$\lambda=0.05$.}
\label{xi5eps}
\end{figure*}

In the case of a $4\times 4$ Hamiltonian, the Chern number can be expressed as
\begin{equation}\label{C2}
C = \frac{1}{2\pi i} \smashoperator{\int_{\rm BZ}}\! d^2k\, \text{Tr} \Big( P^- \big[ \partial_{k_1} P^-, \partial_{k_2} P^- \big] \Big)
\end{equation}
in terms of the projector to the two filled bands $P^- = P_1^- + P_2^-$, where
\begin{equation*}
\begin{split}
P_1^- = \frac{(E_1-H)(E_2-H)(E_{-2}-H)}{(E_1-E_{-1})(E_2-E_{-1})(E_{-2}-E_{-1})},
\\
P_2^- = \frac{(E_1-H)(E_2-H)(E_{-1}-H)}{(E_1-E_{-2})(E_2-E_{-2})(E_{-1}-E_{-2})}.
\end{split}
\end{equation*}
Here $E_{-1}$ and $E_{-2}$ are the energies of the two filled bands and $E_1$ and $E_2$ the energies of the unoccupied bands. Numerical evaluation of the four-band formula (\ref{C2}) requires explicit knowledge of the energy bands $E_i(k)$ of the $4\times4$ Hamiltonian which makes it more costly than two-band expression (\ref{C1}).

\section{Topological phase diagrams}\label{topophase}

In this section, we will present the topological phase diagrams and energy gaps of our model (\ref{h2}) as a function of relevant system parameters in different lattice geometries. The treatment is based on the evaluation of the Chern number by employing Eqs.~(\ref{C1}), (\ref{C2}) and diagonalization of the system in momentum space. Discussion of the detailed structure of the Chern mosaic is postponed to the next section.  

\subsection{Different lattice geometries}

\begin{figure}
\includegraphics[width=1\columnwidth]{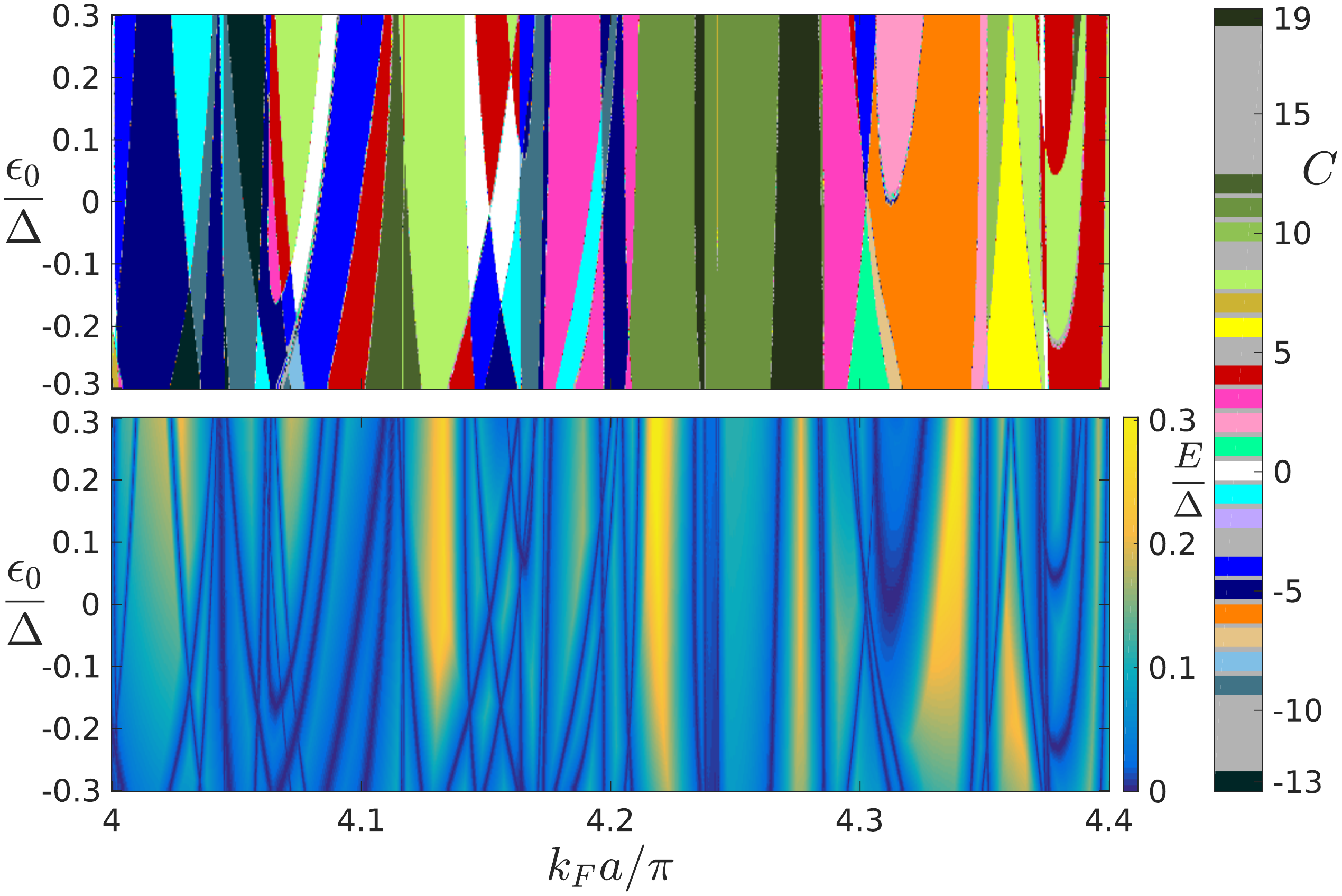} 
\caption{Phase diagram on a square lattice. The same quantities and parameters as in Fig.\ \ref{xi5eps} (a) except $\xi/a=30$.}
\label{xi30eps}
\end{figure}
Below we calculate topological phase diagrams for square, triangular and honeycomb lattices. In Fig.\ \ref{xi5eps} we have plotted phase diagrams classified by different Chern numbers as a function of the decoupled impurity energy $\epsilon_0$ and parameter  $k_Fa$ that controls the hopping between the Shiba sites. Here $a$ denotes the lattice constant of the magnetic lattice. In Fig.\ \ref{xi5eps} we have chosen  a relatively large lattice spacing $\xi/a=5$ which corresponds to separation of a few tens of nanometers for typical superconductors. A phase transition between phases with different Chern numbers is always accompanied by the closing of the energy gap, as revealed by the energy gap diagrams showing the minimum of the positive energy band $\min_kE(k)$. Qualitatively, the phase diagrams corresponding to different lattice geometries depict similar structures, mostly consisting of a topologically nontrivial region with complicated fine structure. Considering that the effective hopping radius $\xi/a=5$ is much larger than the lattice constant, the qualitative similarity between the different geometries is not so surprising. 

\begin{figure*}
\includegraphics[width=0.9\textwidth]{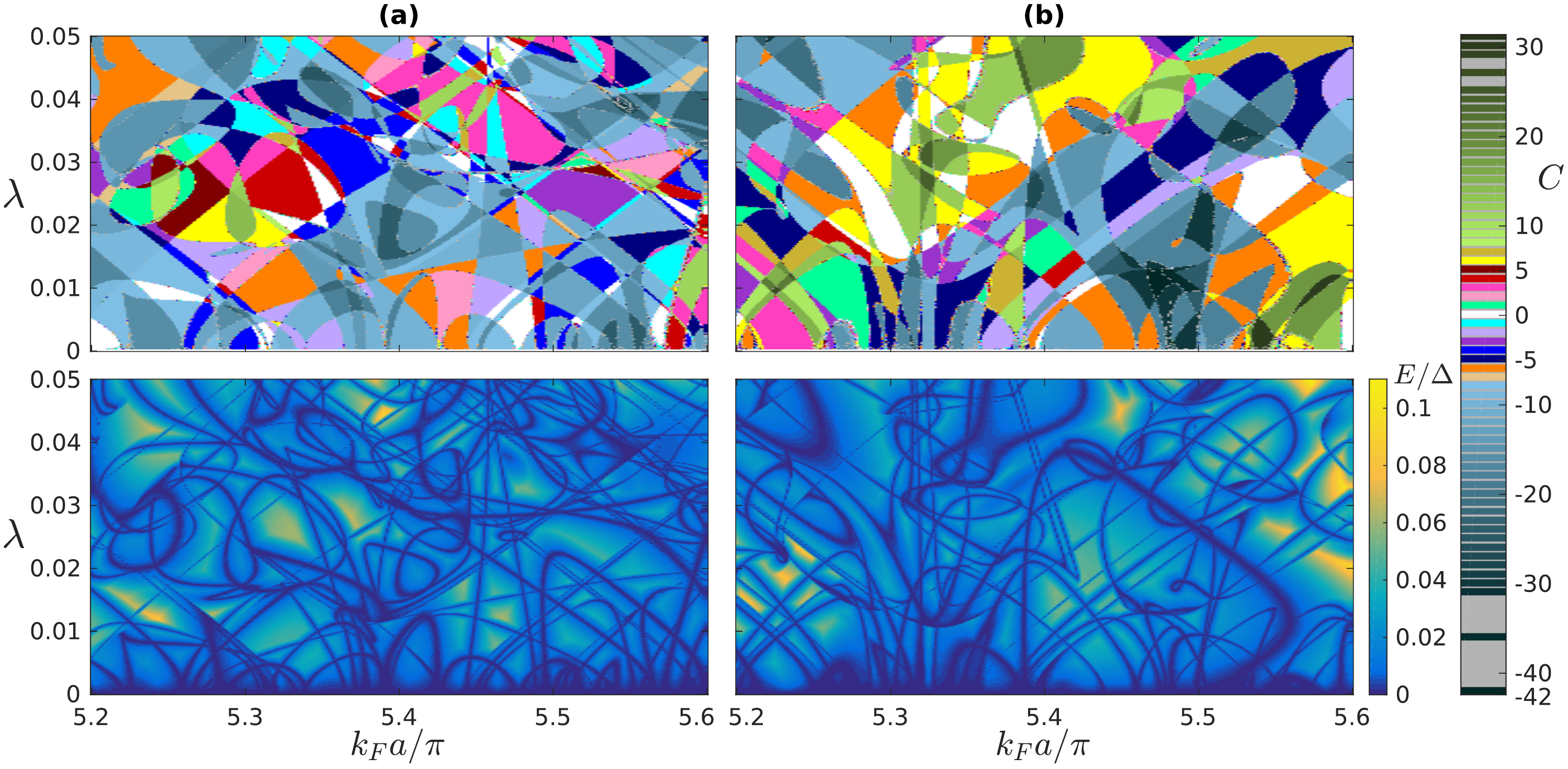} 
\caption{The Chern number (upper panel) and the corresponding energy gap
(lower panel) in terms of $k_Fa$ and the dimensionless SOC strength $\lambda$. The
columns correspond to (a) square and (b) triangular lattices. In all
figures the superconducting coherence length is $\xi/a=10$ and
the uncoupled Shiba state energy $\epsilon_0=0.1\Delta$.}
\label{xi10lam}
\end{figure*}

In Fig.\ \ref{xi30eps} we have depicted the same quantities but chosen a smaller Shiba lattice constant, $\xi/a=30$ than in Fig.\ \ref{xi5eps}. Comparing these two figures, we see that for a larger $\xi$, the topological phase of the system is more sensitive to $k_Fa$, and less sensitive to $\epsilon_0$. This can be qualitatively understood by the fact that a larger hopping radius $\xi/a=30$ will generally increase the hybrization energy as more Shiba sites interact and the single-impurity energy becomes less important. A larger hopping radius will also lead to higher energy gaps. The observed dependence on the lattice spacing is a generic feature of the system.

Figures  \ref{xi5eps} and \ref{xi30eps} together illustrate that  the different lattice geometries generally lead to qualitatively similar topological phase diagram with a rich variety of different Chern number phases. Insensitivity to geometric details is somewhat expected, considering that the typical hopping radius in the model is much larger than the lattice constant.  

\subsection{Role of the Rashba coupling}
In Fig.\ \ref{xi10lam} we show the phase diagram in terms of the hybrization parameter $k_Fa$ and the dimensionless SOC strength $\lambda$ for a fixed ratio $\xi/a=10$. The Rashba coupling is a necessary ingredient to realize non-trivial topological phases. In the absence of the Rashba coupling, $\lambda=0$, the gap function $\Delta_{mn}$ in Eq.~(\ref{h2}) vanishes, implying that the system is either gapless or trivial. In the adopted deep-impurity regime the system at  $\lambda=0$ is generically gapless. A nonzero Rashba coupling will then typically drive the gapless system to a topologically nontrivial state, as indicated in Fig.~\ref{xi10lam}.  It is remarkable how a relatively weak Rashba splitting $\lambda=0.01-0.02$ can translate to  topological energy gaps of the order of $0.1\Delta$ when the decoupled impurity energy $\epsilon_0$ lies near the gap. The Rashba coupling also modifies the effective coherence length but this correction depends on $\lambda^2$ and is insignificant in the realistic parameter regime $\lambda<0.1$. 

\begin{figure*}
\includegraphics[width=0.9\textwidth]{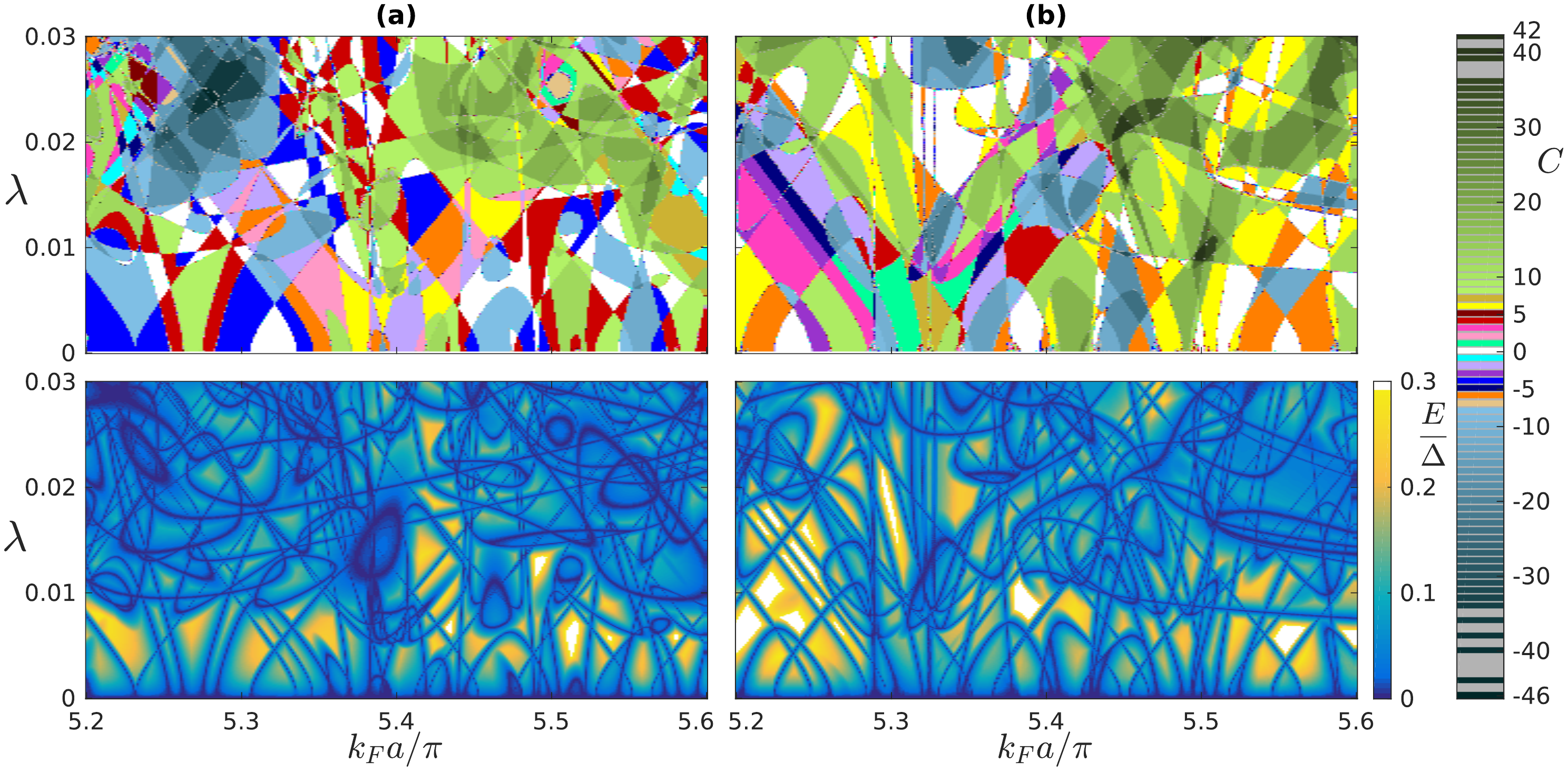}
\caption{Same as Fig.\ \ref{xi10lam} except $\xi/a=30$.}
\label{xi30lam}
\end{figure*}

In Fig.\ \ref{xi30lam} we plot the phase diagram for a ratio $\xi/a=30$. For a larger $\xi$ (alternatively smaller $a$), the effective hopping radius increases and  the increased number of efficiently coupled Shiba states leads to larger excitation gaps. In Fig.\ \ref{xi30lam} we can see that it is possible to have large topological gaps $\sim0.2\Delta$ for small values of the Rashba coupling $\lambda=0.001-0.005$. When the onsite Shiba energy $\epsilon_0$ moves away from the centre, the gaps generally diminish, but it is still striking how weak the Rashba splitting can be and still lead to robust topological states if the Shiba energy lies in the vicinity of the gap center.  

\section{Properties of the Chern mosaic} \label{mosaic}

The physical properties of the model (\ref{h2}) are well-hidden in the complicated analytical structure of the long-range hopping amplitudes that are rapidly oscillating as a function of distance. The emergence of the Chern mosaic is not easily unravelled from the analytical expressions, and we therefore resorted to numerical evaluation of the topological phase diagram. In this section we will extract more detailed information concerning the phase diagram.  

\subsection{Abundance of different phases}
As discussed in the previous section, the phase diagram has a complicated pattern of phase transition lines. The Chern mosaic can be understood as a consequence of the long-range hopping nature of the subgap states.  In the two-band case Eq.~(\ref{C1}) implies a simple interpretation of Chern numbers as the number of times the vector $\hat{\vr{d}}(\vec{k})=\vec{d}/|\vec{d}|$ covers the unit sphere as $k$ takes values in a Brillouin zone. In a square lattice, a hopping between the $n$th neighbours in $x$ and $y$ direction gives rise to terms proportional to $\cos(nk_{x/y}a)$ in $d_z(\vec{k})$, and $\sin(nk_{x/y}a)$ in $d_{x/y}(\vec{k})$, that oscillate more rapidly with increasing $n$. Thus, the number of times $\hat{\vr{d}}$ may cover the unit sphere will generally increase with hopping distance $n$. The asymptotic forms for the Bessel and Struve function indicate that the $n$th hopping terms scale as $[|\Delta|/(k_Fa)^{1/2}](e^{-an/\xi}/n^{1/2})$ for large $n$, so the decay is very slow for the hopping range $n<\xi/a$. On top of the smooth envelope, the $n$th hopping terms are proportional to  $\sin {nk^\pm_Fa}$ or $\cos {nk^\pm_Fa}$. These oscillations cause the phase diagrams to vary rapidly as a function of $k_Fa$. The Chern mosaic structure thus emerges from the effective competition of at least $\mathcal{O}(\xi/a)$ different hopping terms.   

\begin{figure}
\includegraphics[width=0.49\columnwidth]{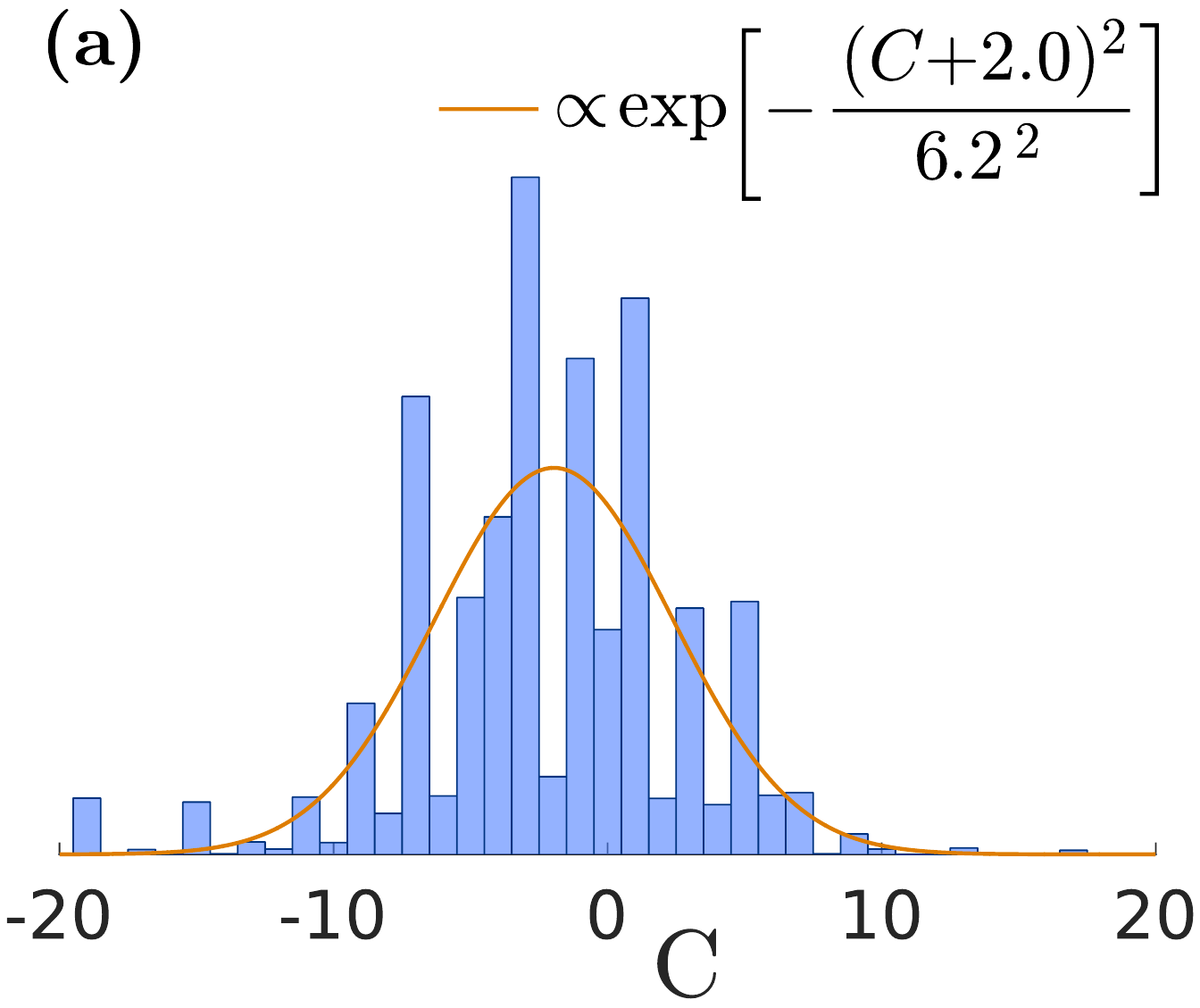} 
\includegraphics[width=0.49\columnwidth]{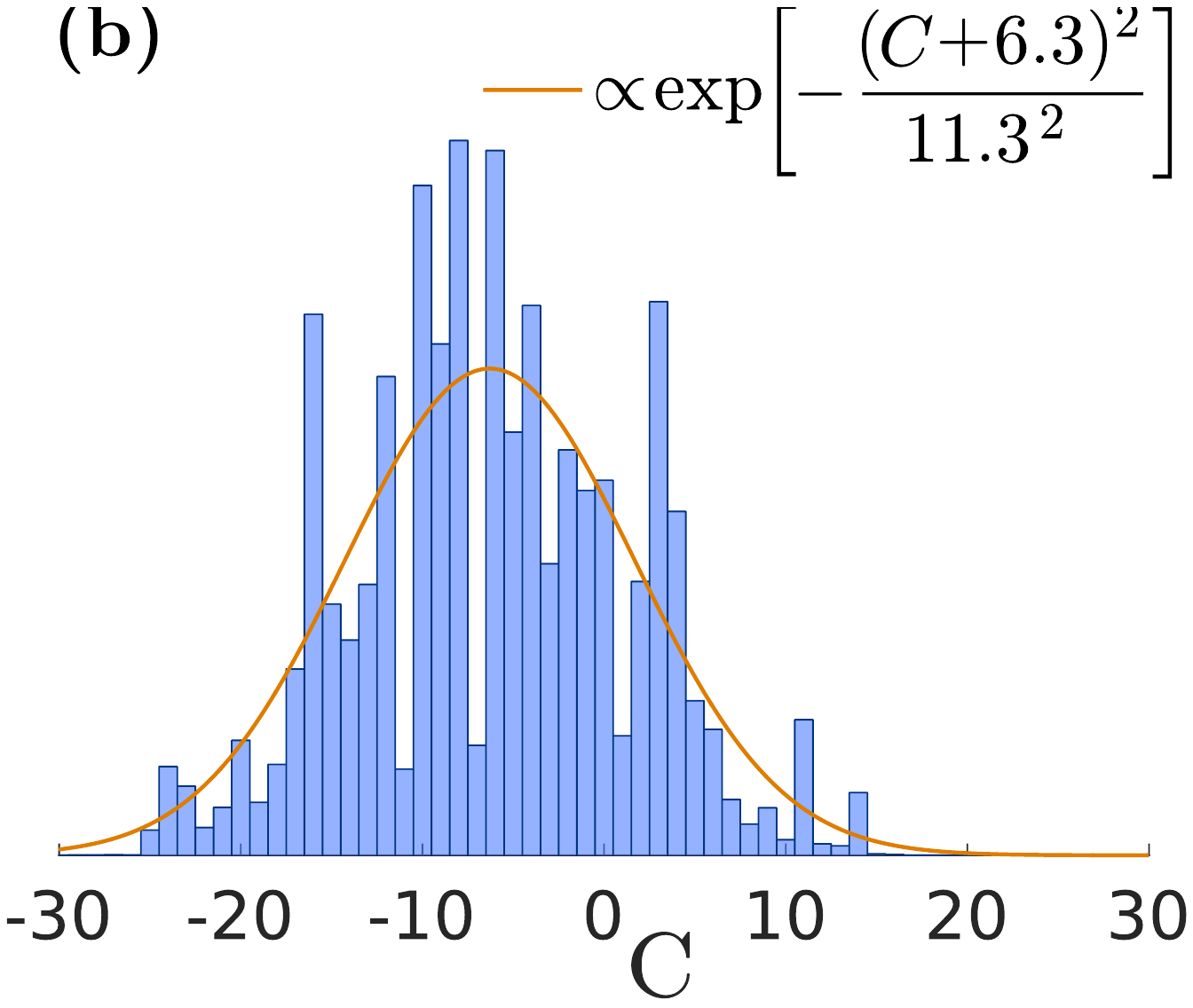} 
\caption{Relative areas of different phases calculated from the phase diagrams in
Figs.\ \ref{xi5eps} (a) (left) and \ref{xi10lam} (a) (right). We have discarded those
Chern numbers which are not exactly quantized to integers due to numerical
inaccuracy near the phase boundaries. We have included sample points with
a Chern number $C$ satisfying $|C-q|<0.1$ for an integer $q$.}
\label{Cdist}
\end{figure}

In Fig.~\ref{Cdist} we have plotted the relative areas of different phases for two different phase  diagrams presented in Figs.~\ref{xi5eps} and \ref{xi10lam}. For a finite $k_Fa$ window the center of the histogram does not coincide with $C=0$. The width of the fitted Gaussian serving as a guide to the eye  indicates that the width of the distribution is $\mathcal{O}(\xi/a)$ as expected on theoretical grounds.

\subsection{Energy gap vs. Chern numbers}
From the results presented above we have seen that the energy gaps protecting the topologically nontrivial phases in various circumstances can be of the order 0.1$\Delta$ corresponding to a temperature 1K.  It is interesting to see how the size of energy gaps depends on the magnitude of the Chern number. Due to the complicated form of the long-range hopping model, this question is difficult to address  analytically. On theoretical grounds we expect the gaps to decrease as function of the Chern number. This is because increasing hopping radius generates higher Chern numbers, but the hopping amplitudes themselves are suppressed as $e^{-\frac{an}{\xi}}/\sqrt{n}$ for the $n$th neighbours. 

\begin{figure}
\includegraphics[width=0.49\columnwidth]{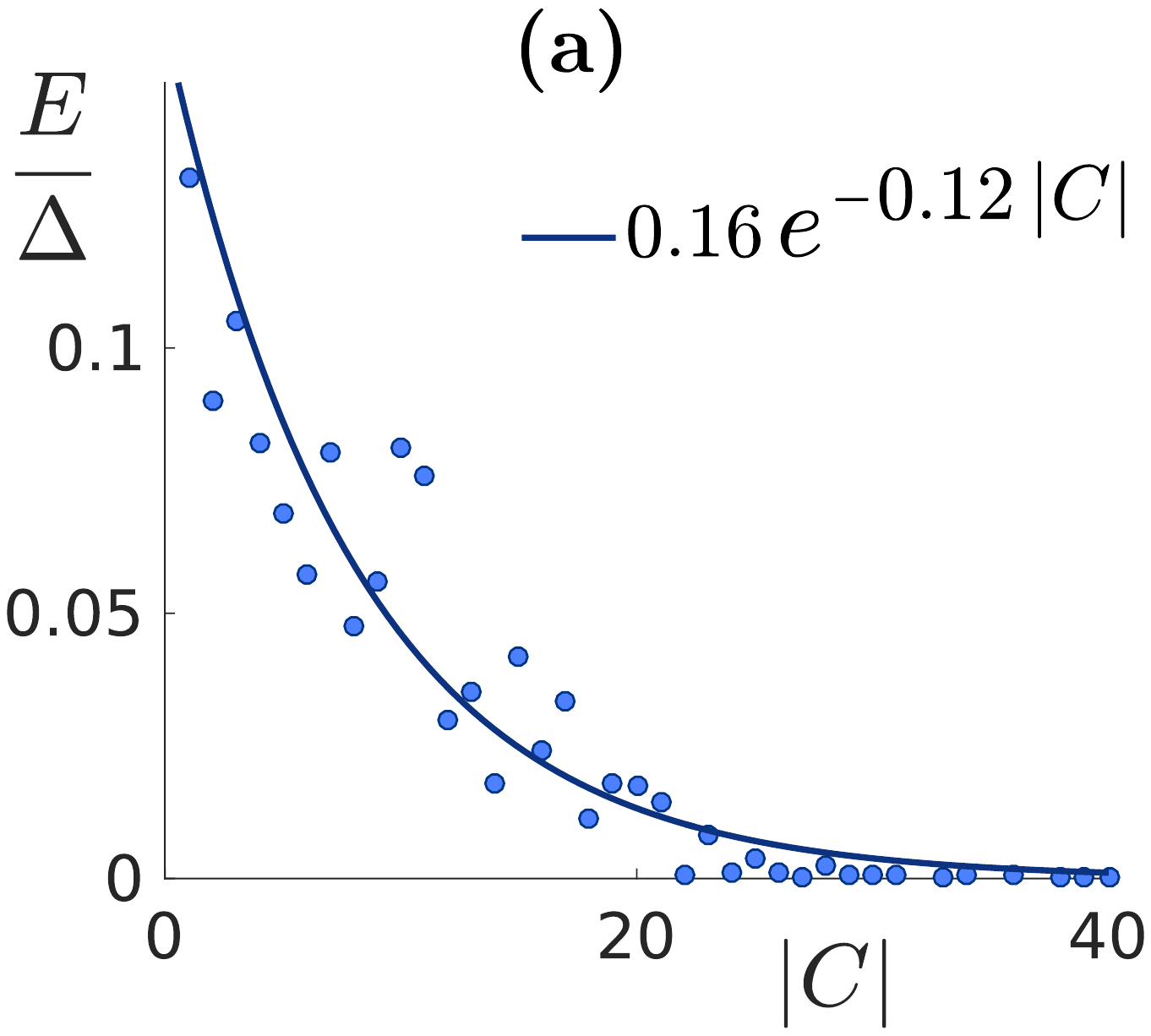} 
\includegraphics[width=0.49\columnwidth]{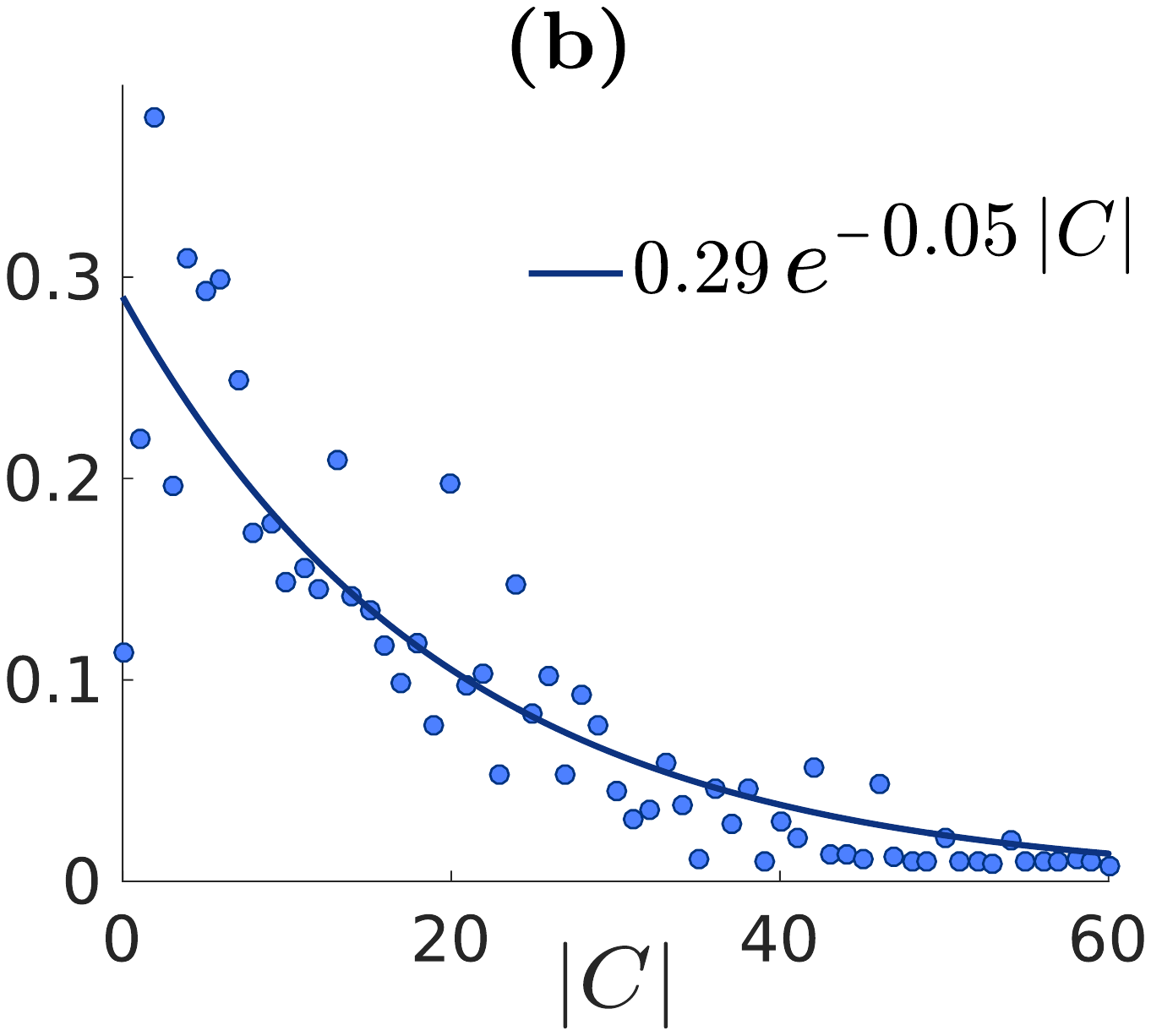} 
\caption{The dependence of the energy gap on the Chern number $C$. We plot
the maximum gap found for each $|C|$ for (a) $\xi/a=5$, $\lambda=0.05$ and (b)
$\xi/a=30$, $\lambda=0.02$. We have sampled the system over a parameter window
$k_Fa/\pi=4\dots8$, $\epsilon_0/\Delta=-0.3\dots0.3$.}
\label{fitfig}
\end{figure}

In Fig.\ \ref{fitfig} (a) we plot the maximum energy  gaps as a function of the magnitude of the Chern number on a square lattice corresponding to fraction $\xi/a=5$. The data has been collected from a fixed $k_Fa, \epsilon_0$ parameter window. The exponential fitting, motivated by the exponential suppression of the hopping and included for a guide to the eye, captures the decreasing gap trend reasonably well. In Fig.\ \ref{fitfig} (b) we plot the gap as a function of the Chern number for a larger coherence length $\xi/a=30$. As expected, the decreasing trend is now significantly slower.   


Since the occurrence of large Chern numbers $\gg \xi/a$ is suppressed, the conclusions regarding their properties are limited.  While the illustrated results correspond to snapshots from a specific parameter window, the exponential suppression of the gaps as a function of the Chern number seems reasonable description.

\section{Discussion and outlook} \label{summary}

In this work we have studied exotic chiral superconductivity in superconducting films with a Rashba spin-orbit coupling decorated by lattices of magnetic adatoms. The rich topological properties and high Chern numbers emerge generically in  different lattice geometries when the magnetic moments are ferromagnetically ordered.  The magnetic ordering is consequence of various different mechanisms in real systems and is difficult to predict theoretically. The electron-mediated indirect RKKY coupling and spin-orbit coupling will generally favor spiral magnetic order.\cite{brau1,brau2,klin,schecter} Perpendicular ferromagnetic alignment is favored by crystal field anisotropy on the surface\cite{heimes2} and could possibly  be affected by moderate external magnetic fields that do not destroy superconductivity. In the case where magnetic moments form ferromagnetic domains with opposite magnetization perpendicular to the surface, the different domains support antichiral superconductivity. When magnetization is inverted, the factor $x_{mn}-iy_{mn}$ in Eq.~(\ref{h2})  changes to $x_{mn}+iy_{mn}$ and the Chern number changes its sign as required by time-reversal transformation. Thus, different ferromagnetic domains are separated by gapless edge states. 

\begin{figure}
\includegraphics[width=0.7\columnwidth]{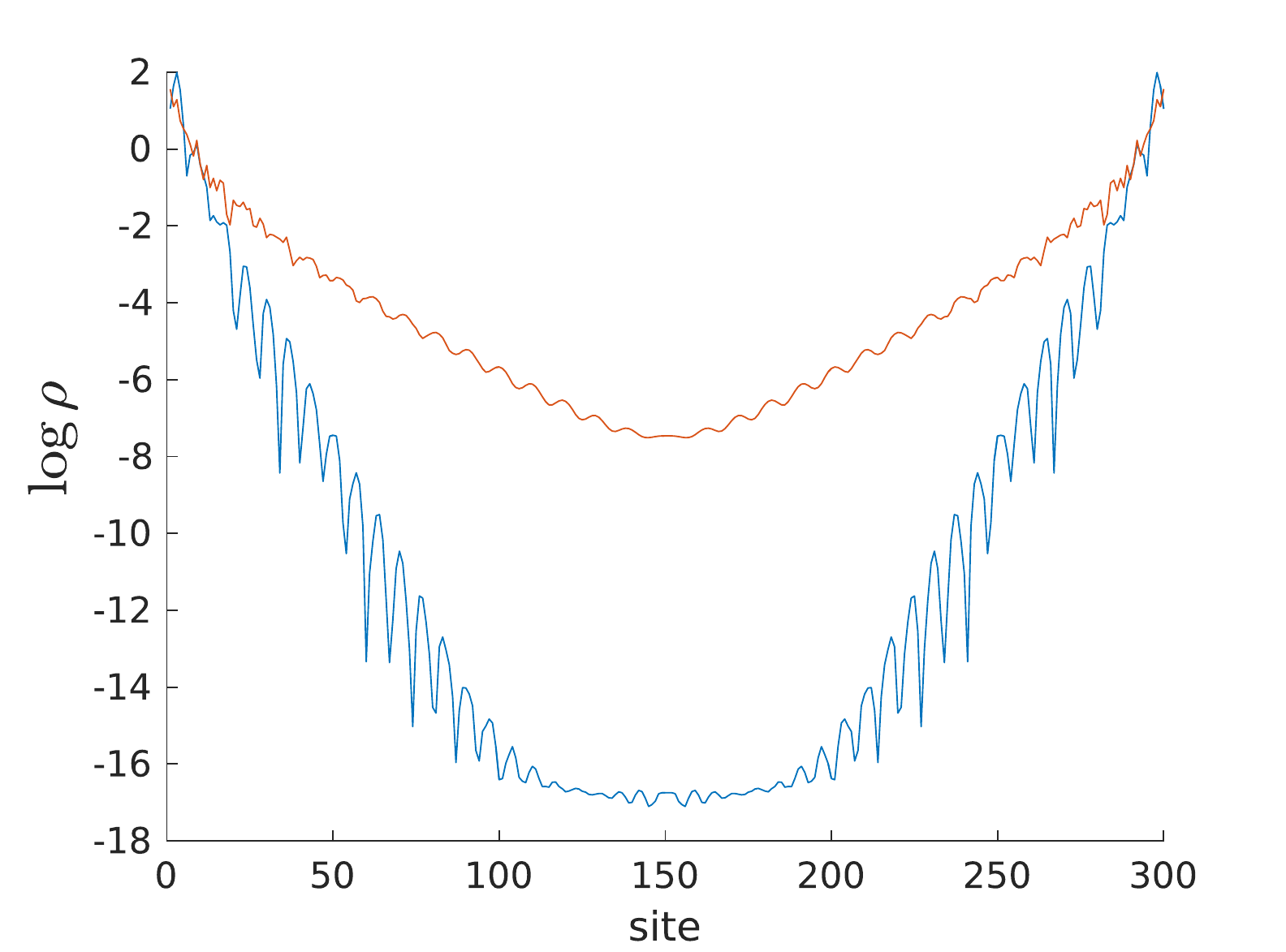} 
\caption{Logarithm of local density of states $\rho$ (arb.\ units) at $E=0$ (averaged in the window $E/\Delta\in[-0.01,0.01]$) as a function of distance from one edge on a finite strip of 300 lattice sites. LDOS arises from the chiral edge states propagating along the edge. The red curve corresponds to a $C=-9$ state with parameters $\xi/a=5$,
$k_Fa/\pi=9.83$, $\epsilon_0=0.18\Delta$, $\lambda=0.05$  and the blue curve represents a $C=3$ state with parameters $\xi/a=10$, $k_Fa/\pi=4.9$, $\epsilon_0=-0.22\Delta$, $\lambda=0.05$.   }
\label{dos}
\end{figure}

As discussed in Ref.~\onlinecite{ront1}, signatures of the edge states can be accessed through STM measurements by probing the local density of states (LDOS) at subgap energies. All states below the bulk energy gap must arise from the edge states. As shown in Ref.~\onlinecite{ront1}, these states show up as enhanced LDOS near the edge of magnetic lattice.  In Fig.~\ref{dos} we have plotted the spatial decay of the midgap LDOS for states with Chern numbers $C=-9$ and $C=3$ with energy gaps $E=0.07\Delta$ and $E=0.05\Delta$. Both states show signatures of exponential decay with some modifications. Pure exponential decay is not expected in the studied long-range hopping model, as discussed in the treatment of 1d version and Majorana bound states in Ref.~\onlinecite{poyh2}. While an analytical expression for the penetration depth is not known, Fig.~\ref{dos} implies that the states are reasonably rapidly decaying into the bulk. 

As a physical realization one could consider for example Pb or Nb surfaces while possible magnetic adatoms  include Fe, Co, Cr and Mn. In a recent experiment\cite{menard}, Shiba states were observed in the vicinity of Fe impurities on 2H-NbSe$_2$, a layered transition metal dichalcogenide that becomes an anisotropic $s$-wave superconductor below 7.2 K. This experimental observation demonstrates how quasi-2d surface superconductivity supports long-range Shiba wavefunctions. In a quasi-2d geometry as assumed in our work, theory predicts a slow $1/\sqrt{r}$ decay of Shiba states before the onset of the exponential decay  at $r\sim\xi$, in contrast to the $1/r$ decay in a 3d bulk. Experiments in Ref.~\onlinecite{menard} showed that spatial disturbances due to the Shiba states can extend tens of nanometers away from the impurity, which is an order of magnitude longer than previous observations. Since complex topological properties arise precisely from the long-range nature of the Shiba states, the recent experimental observation is very interesting in this context.

Artificial lattices of the order of a few hundred atoms could be constructed by STM methods but self-assembly techniques would probably be required for larger systems. The size of the system which displays clear signatures of the topological edge states depends on the system parameters, but a few hundred atoms should prove sufficient.\cite{ront1} The Rashba spin-orbit coupling is a crucial ingredient in achieving chiral superconductivity with a ferromagnetic texture. As we have seen above, a Rashba splitting of the order of $|k_F^+-k_F^-|/k_F\sim~0.01$ may already lead to energy gaps of the order of $0.1\Delta$ under favorable circumstances. Therefore, the magnitude of the required Rashba coupling could be obtained, for example, in Pb films.\cite{dil} 

In this work we considered ferromagnetic textures which, together with the Rashba coupling in the bulk electrons, were responsible for the nontrivial topological superconductivity. However, this is not the sole route to nontrivial phases.  As in 1d chains, it is possible to achieve a topologically nontrivial superconductor in 2d lattices with helical ordering. In that case, the Rashba coupling is not necessary. This type of system was analyzed in Ref.~\onlinecite{li1} within a simplified nearest neighbour toy model that ignores the microscopic structure of Shiba states. Helical textures could also be studied within the theoretical framework of Eq.~(\ref{h1})  by specifying a helical texture in Eq.~(\ref{hmn_Dmn_general}). Thus the formalism employed in our work may be applied to future work in this direction.  

 Phases with high Chern numbers and rich topological properties in the studied system arise from the adatom patterning on a trivial superconductor. A conceptually similar approach could be applied to other gapped and non-superconducting systems where impurities hybridize and form subgap bands.\cite{kimme} These bands can support high topological numbers and provide a route to engineer topologically nontrivial states different from the parent state in the absence of adatoms. If successful, this program could pave the way toward increasingly complex man-made topological phases. 

\acknowledgements
The authors acknowledge Alex Weststr\"om and Kim P\"oyh\"onen for valuable discussions and comments on the manuscript and the computational resources provided by Aalto Science-IT project. T. O. acknowledges the Academy of Finland and J. R. the Finnish Cultural Foundation for support.

\end{document}